\documentclass[usenatbib,useAMS]{mnras}
\RequirePackage{fix-cm}
\usepackage{txfonts}
\usepackage{ae,aecompl}
\usepackage{graphicx}	
\usepackage{aas_macros}
\usepackage{epsfig,graphicx}
\usepackage{longtable}
\usepackage{amsfonts}
\usepackage{amssymb}
\usepackage{wasysym}
\usepackage{color}
\usepackage{tabularx}
\usepackage{aas_macros}
\usepackage{comment}
\usepackage[farskip=0pt]{subfig}

\newcommand{\plck}{\emph{Planck}\,\,}
\newcommand{\xmm}{\emph{XMM-Newton}\,\,}

\newcommand{\myclus}{PLCK\,G200.9$-$28.2\,\,}
\newcommand{\mjyb}{mJy beam$^{-1}$}

\title[Relic in PLCKG200.9-28.2]{Discovery of a radio relic in the low mass,  
merging galaxy cluster \myclus}
\author[R. Kale et al.]
{Ruta Kale,$^{1}$\thanks{E-mail:ruta@ncra.tifr.res.in}
Daniel R. Wik,$^{2,3,4}$
Simona Giacintucci,$^{5}$ 
Tiziana Venturi,$^{6}$
\newauthor Gianfranco Brunetti,$^{6}$
Rossella Cassano,$^{6}$ 
 Daniele Dallacasa$^{6,7}$
 and 
 \newauthor Francesco de Gasperin$^8$\\
{$^{1}$National Centre for Radio Astrophysics, Tata Institute of 
Fundamental 
Research, Post Bag 3, Pune 411007, India}\\
{$^{2}$NASA Goddard Space Flight Centre, Code 662, Greenbelt, MD 20771, USA}\\
{$^{3}$The Johns Hopkins University, Homewood Campus, Baltimore, MD 21218, 
USA}\\
{$^{4}$Department of Physics and Astronomy, University of Utah, 115 
South 1400 East, Salt Lake City, UT 84112, USA}\\
{$^{5}$Naval Research Laboratory, 4555 Overlook Avenue SW, Code 
7213, 
Washington, DC 20375, USA}\\
{$^{6}$INAF-Istituto di Radioastronomia, via Gobetti 101, 40129 Bologna, Italy 
}\\
{$^{7}$Dipartimento di Fisica e Astronomia, Universita di Bologna, via
Ranzani 1, 40127 Bologna, Italy}\\
{$^8$ Leiden Observatory, Leiden University, P.O.Box 9513, NL-2300 RA Leiden, 
The Netherlands}
}

\date{Accepted XXX. Received YYY; in original form ZZZ}

\pubyear{2017}

\begin{document}
\label{firstpage}
\pagerange{\pageref{firstpage}--\pageref{lastpage}}
\maketitle

\begin{abstract}
Radio relics at the peripheries of galaxy clusters are tracers of 
the elusive cluster merger shocks. We report the discovery of a single radio 
relic in the galaxy cluster \myclus ($z=0.22$, $M_{500} = 2.7\pm0.2 
\times 10^{14} M_{\odot}$)
using the Giant Metrewave Radio Telescope at 235 and 610 MHz and the Karl G. 
Jansky Very Large Array at  1500 MHz. 
The relic has a size of $\sim 1 \times 0.28$ 
Mpc, an arc-like morphology and is located at 0.9 Mpc from 
the X-ray brightness peak in the cluster. The integrated spectral index of the 
relic is $1.21\pm0.15$. The spectral index map between 235 and 610 MHz shows 
steepening from the outer to the inner edge of the relic in line with the 
expectation from a cluster merger shock. Under the assumption of diffusive 
shock acceleration, the radio spectral index implies a Mach number of 
$3.3\pm1.8$ for the shock.
The analysis of archival \xmm data shows that \myclus consists of a 
northern brighter sub-cluster, and a southern sub-cluster in a state of 
merger. This cluster has the lowest mass among the clusters hosting single 
radio relics. The position of the \plck Sunyaev Ze'ldovich effect 
 in this cluster is offset by 700 kpc from the X-ray peak in the direction of 
the radio relic, suggests a physical origin for the offset. Such large  
offsets in low mass clusters can be a useful tool to select disturbed clusters 
and to study the state of merger.
\end{abstract}


\section{Introduction}
Radio relics are diffuse, elongated, radio sources of synchrotron origin, 
occurring at the peripheries of galaxy clusters 
 in the form of spectacular single or double symmetric arcs around cluster 
centres \citep[e.g.][]{bag06,gia08,wee10,gas14}. 
 These are exclusively found in clusters undergoing mergers and have been 
proposed to originate in acceleration/ re-acceleration at 
 merger-shocks \citep[see][for a review]{bru14}. The scenario is supported by 
several observational facts:
The radio spectral index distribution across relics typically shows a flat 
to steep trend from the outer to the inner edge 
\citep[e.g.][]{oru07,wee10,wee12,kal12,bon12,str13,gas15}. Relics have been 
found to be polarized ($20-30$ per cent), indicating 
underlying ordered magnetic fields \citep[e.g.][]{wee10,kal12,gas15}. 

Diffusive shock acceleration (DSA) of electrons at shocks driven by merging sub-clusters 
 can explain the power-law spectra of radio relics \citep[e. g.][]{ens98,hoe07,kan05,kan12}. 
However, DSA has low acceleration 
efficiency in the low Mach number shocks ($M < 4$) typical in galaxy clusters 
\citep[e. g.][]{mar10} and thus, is not sufficient to explain some of the 
observed large radio relics \citep{bru14}. 
The observed spectral curvatures and 
widths of a few radio relics cannot be explained under standard DSA 
\citep[e. g.][]{str13,ogr14,tra14,str16}. The contamination due to the
Sunyaev-Ze'ldovich (SZ) effect can explain the 
spectral steepening in some relics \citep{basu16}. 

Recently, using the particle-in-cell simulations of weak collisionless 
shocks, Shock Drift Acceleration (SDA) has been proposed as a pre-heating 
mechanism \citep[e. g.][]{matsu11,guo14}. This can alleviate the issue of low 
acceleration efficiency with the standard DSA providing a first stage of 
re-acceleration. The remnants of radio galaxies  available at the location of 
the shock are also plausible sources of seed relativistic electrons to DSA 
\citep[e. g.][]{mar05,kang15}. However the role of magnetic 
field, the efficiency of acceleration, and injection of seed relativistic 
electrons in the shock is still not well understood
  \citep[][]{bru14}.

Radio relics are known to occur as single or double arcs at cluster 
peripheries. Double relics are special systems where the associated cluster 
merger axis is nearly in the plane of the sky 
and allows the detection of the shock properties in radio and X-rays. For 
the sample of known double radio relics a scaling relation between the 1.4 GHz 
radio power of the relic and the host cluster mass has been reported by 
\citet{gas14}. It was also shown that the single relics show a large scatter 
around this scaling relation. Galaxy clusters with single arc-like relics may 
be cluster mergers with a geometry unfavourable to allow 
the view of the second relic or are intrinsically 
systems with merger parameters that did not result in the generation of the 
second relic. A large sample of relics is needed to understand the connection 
between mergers and generation of relics.

The number of known relics associated with merger shocks is still small. 
Searches for relics have been  
carried out in all sky radio surveys such as the NRAO VLA Sky Survey 
\citep[NVSS][]{con98, gio00}, the Westerbork Northern Sky Survey 
\citep[WENSS][]{wenss,kem01} and the VLA Low Frequency Sky Survey 
\citep[VLSS][]{vlss,wee09} in the directions 
of galaxy clusters. Radio relics have also been discovered using observations 
with the Green Bank Telescope in the directions of clusters \citep{gbtrelics}.
In the Extended GMRT Radio Halo Survey sample 
that consisted of clusters with X-ray luminosity $>5\times10^{44}$ erg s$^{-1}$ 
in the redshift range of 0.2 - 0.4, the radio relics were found in $\sim 5$ per 
cent of the clusters \citep{kal15}.

Galaxy clusters with indications of an ongoing merger are potential 
candidates as hosts of radio relics. Recently the \plck satellite, using the 
SZ decrement, has detected 947 clusters of which 214 
clusters are new \citep{2015A&A...581A..14P}. 
The newly discovered clusters have lower luminosities and flatter density profiles as compared to the 
clusters already known from X-ray observations \citep[][hereafter 
PC12]{2012A&A...543A.102P}.
These properties indicate ongoing mergers and therefore the new \plck clusters are potential candidates for 
the search of radio relics in radio surveys.

Radio relics
have been reported in the new \plck clusters since the \plck Early 
SZ catalogue of galaxy clusters \citep[][]{bag11,gas14,gas15}. In this paper,  
we report the discovery of a radio relic at the periphery of a new \plck 
cluster, PLCKG200.9-28.2, using radio observations from the Giant Metrewave 
Radio 
Telescope (GMRT) and the Karl G. Jansky Very Large array (VLA). The paper is 
organized as follows: The galaxy cluster \myclus is introduced in 
Sec.~\ref{clus}. The radio data analysis and the images are presented in 
Secs.~\ref{obs} and ~\ref{images}. 
The analysis of \xmm data towards this cluster is presented in 
Sec.~\ref{xrayobs}. The origin of the radio relic is discussed in 
Sec.~\ref{discussion}. Conclusions are 
presented in Sec.~\ref{conc}.

We adopt $\Lambda$CDM cosmology with $H_0 = 71$ km s$^{-1}$ Mpc$^{-1}$, 
$\Omega_\Lambda=0.73$ and 
$\Omega_m = 0.27$ in this work. This implies a linear scale of 3.52 kpc per 
arcsec at the redshift of the cluster 
PLCKG200.9-28.2.

\section{PLCK G200.9-28.2}\label{clus}
\myclus was a candidate cluster in the \plck satellite Early 
Cluster catalogue \citep[][hereafter PC11]{2011A&A...536A...8P} that was 
subsequently confirmed with the \xmm observations in X-rays (PC12). It is 
a cluster at a redshift of 0.22 with mass $M_{500} = (2.7\pm0.2) \times 
10^{14} 
M_{\odot}$, X-ray luminosity $L_{500,[0.1-2.4]keV} = (0.99\pm0.04) 
\times10^{44}$ 
erg s$^{-1}$, and an 
average temperature of $4.5$ keV (Table ~\ref{clusprop}). This cluster has a 
large offset of $3.4'$ between the X-ray and the SZ position, which is above 
the median offset of $1.6'$ \citep{2011A&A...536A...8P,2011A&A...536A...9P} for 
other clusters in the \plck catalogue and 
the \plck reconstruction uncertainty that peaks at $2'$ (PC12). 
An extended radio source at the edge of this cluster was noticed in the NVSS  
and was identified as a candidate radio relic \citep{kal13}. In this work, we 
present deep radio observations and a re-analysis of the X-ray data that allow 
us to study the nature of the radio relic and the host cluster.

\begin{table}
\centering
\caption[]{\label{clusprop}Properties of \myclus (PC12).}
\begin{tabular}{ll}
\hline\noalign{\smallskip}
RA$_{\rm J2000}$ & 04h50m20.9s \\
\smallskip
DEC$_{\rm J2000}$ &-02$^{\circ}$56$^\prime$57.6$^{\prime\prime}$ \\
\smallskip
Redshift & 0.22\\
\smallskip
$L_{500,[0.1-2.4]keV}$ & $(0.99\pm0.04) \times10^{44}$ erg s$^{-1}$\\
\smallskip
$M_{500}$ & $(2.7\pm0.2) \times 10^{14}$ 
M$_{\odot}$\\
\noalign{\smallskip}
\hline\noalign{\smallskip}
\end{tabular}
\end{table}

\begin{table*}
\begin{center}
\caption[]{\label{obstab} Summary of the radio observations.}
\begin{tabular}{cccccccc}
\hline\noalign{\smallskip}
Telescope&Date & Freq.& BW   & Freq. res.&Time &Beam&$\sigma_{\mathrm{rms}}$ \\
         & &MHz  & MHz &kHz/channel& min. & $''\times''$, $^\circ$&mJy beam$^{-1}$\\
\hline\noalign{\smallskip}
GMRT     &31 Dec. 2012 &235 & 8 & 130&390&$14.7\times12.8$, $50.4$  &0.39\\
GMRT     &31 Dec. 2012 &610 & 32 &130& 390& $5.6\times4.7$, $59.4$  &0.045 \\
VLA     &04 Jul. 2013 &1500 & 1000& 25& 90&$14.5\times10.0$, $33.7$ &0.060\\
\hline\noalign{\smallskip}
\hline\noalign{\smallskip}
\end{tabular}
\end{center}
\end{table*}

\begin{figure*}
\subfloat[GMRT 610 MHz]{\includegraphics[trim=0.3cm 
0.3cm 0.3cm 1.8cm,clip,height=8.2cm]{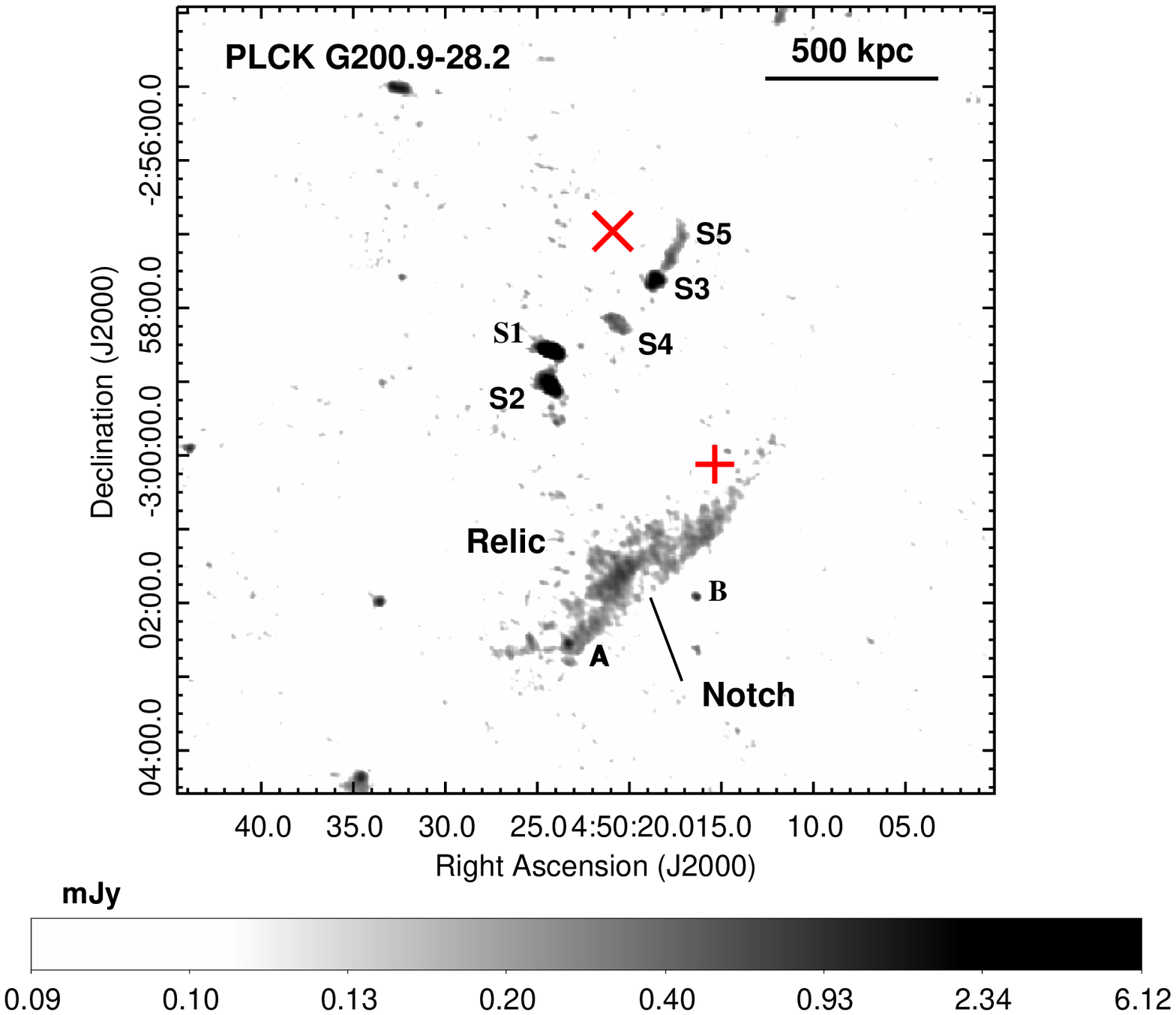}\label{gmrt610}}
\hspace{0.5cm}
\subfloat[GMRT 610 MHz contours on DSS R-band]{\includegraphics[trim=0.9cm 
0.6cm 2.0cm 2.1cm,clip,height=8.2cm]{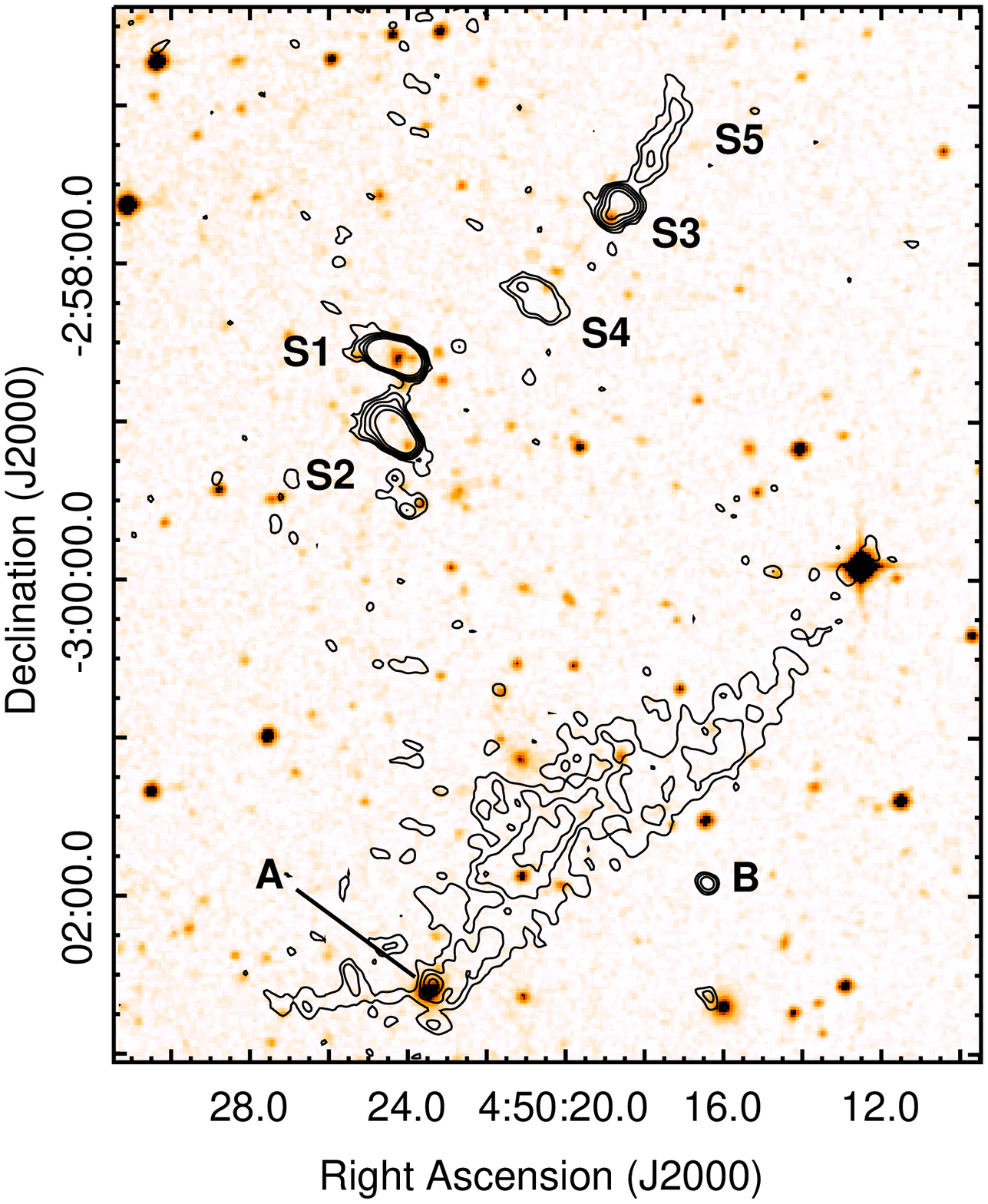}\label{gmrtopt}}
\caption{(a) -- GMRT 610 MHz image is shown in grey scale. The 610 MHz 
image has a 
resolution of $5.6''\times4.7''$, position angle $59.4^{\circ}$ and rms noise, 
$\sigma_{\mathrm{rms}}=$ 0.045 \mjyb.  The discrete sources in the central 
region are 
labelled S1 -- S5 and 
those near the extended source labelled `Relic' are A and B. The symbols 
`$\times$' and `$+$' mark the positions of 
the cluster according to the \xmm and the \plck measurements, 
respectively. (b) -- Digitized Sky Survey R-band image of the region of the 
labelled sources is shown in grey scale with the GMRT 610 MHz contours 
overlaid. The contour levels are $3\sigma_{\mathrm{rms}}\times[\pm1, 2, 4, 
...]$ \mjyb.
 }\label{fig1}
\end{figure*}

\begin{figure*}
\subfloat[GMRT 235 MHz]{\includegraphics[trim=0.5cm 
0.5cm 1.8cm 1.4cm,clip,height=7.8cm]{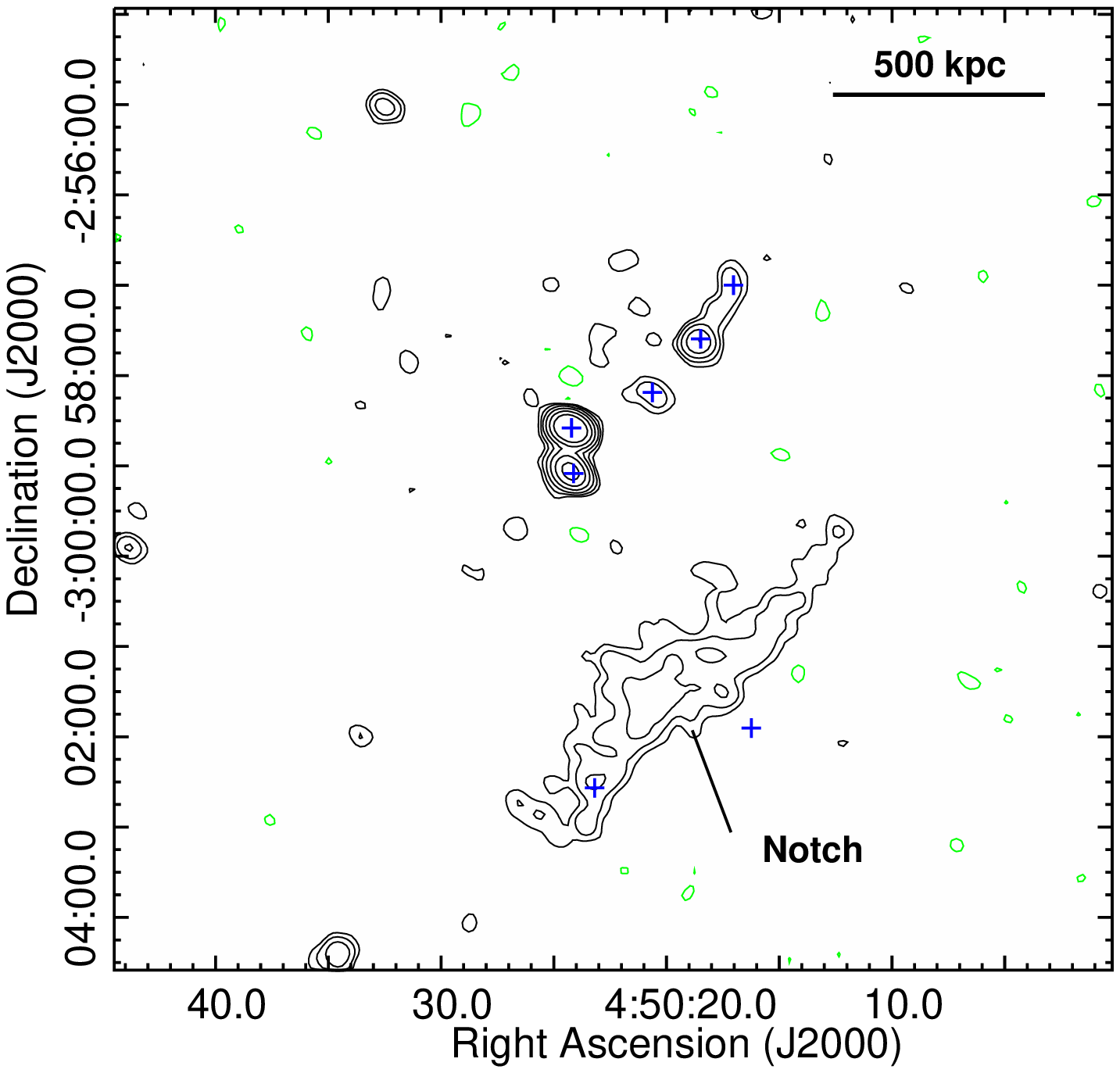}\label{gmrt235}}
\subfloat[VLA 1500 MHz]{\includegraphics[trim=0.5cm 
0.5cm 1.8cm 1.4cm,clip,height=7.8cm]{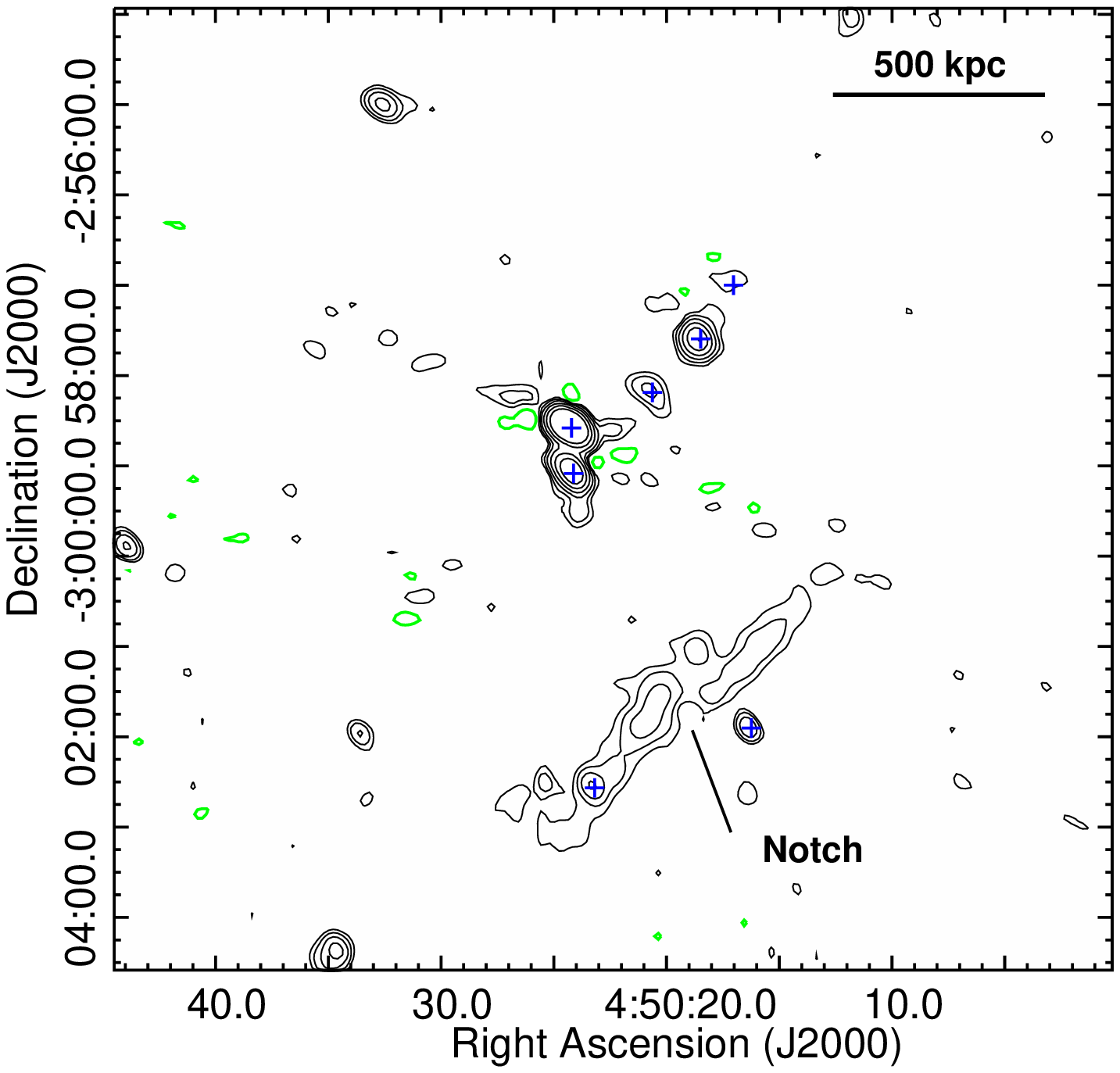}\label{jvla}}
\caption{(a) -- GMRT 235 MHz image 
with $\sigma_{\mathrm{rms}}=$ 0.39 \mjyb is and resolution of 
$14.7''\times12.8''$, 
position angle $57.6^{\circ}$  
is shown in contours.  (b) -- VLA 1.5 GHz image with a resolution of  
$14.5''\times10.0''$, position angle
$33.7^{\circ}$ and $\sigma_{\mathrm{rms}}= 0.06$ \mjyb is shown in contours. 
In panels a and b, the black are positive and the green are negative contour 
levels. The plus signs mark the discrete sources labelled in 
Fig.~\ref{fig1}.\label{fig2}}
\end{figure*}

\begin{center}
\begin{table*}
\caption[]{\label{srctab}Radio properties of the discrete radio sources in the 
\myclus \,field. 
}
\begin{tabular}{ccccccccccccc}
\hline\noalign{\smallskip}
Source & $RA_{J2000}$ & $DEC_{J2000}$ & $S_{235MHz}$        & $S_{610MHz}$  &  
$S_{1500MHz}$  & Spectral index\\
     &	hh mm ss    &$\circ$ $\prime$ $\prime\prime$		    & mJy        
      & mJy        &    mJy    &  $\alpha$   \\
\hline\noalign{\smallskip}
S1   & 04 50 24.57  & -02 58 34.6   & $ 149\pm15 $   & $ 89 \pm 5   $ & 
$49.4\pm4.9$ & $0.63\pm0.04$  \\
S2   & 04 50 24.10  & -02 59 07.6   & $  70\pm7  $   & $ 39 \pm 2   $ & 
$18.5\pm1.9$  & $0.79\pm0.06$ \\
S3   & 04 50 18.56  & -02 57 36.6   & $  25\pm3  $   & $ 14 \pm 1   $ & 
$8.4\pm1.0 $  & $0.58\pm0.01$ \\
S4   & 04 50 20.69  & -02 58 15.6   & $  68\pm7  $   & $ 35 \pm 2   $ & 
$1.7\pm0.2 $ & $0.7\pm0.1,3.4\pm0.1^{\dag}$\\
S5   & 04 50 17.82  & -02 57 17.6   & $  11\pm1  $   & $ 4.9 \pm 0.3$ & 
$0.6\pm0.1$  &  $0.9\pm0.1,2.3\pm0.2^{\dag}$ \\
A    & 04 50 23.36  & -03 02 33.6   & -              & $ 3.2 \pm 0.2 $ & 
$1.5\pm0.2$ & $0.8\pm0.2$  \\           
B    & 04 50 16.42  & -03 01 54.6   & -              & $ 0.96 \pm 0.10 $ 
&$1.5\pm0.2$  & $-0.5\pm0.2$  \\
\hline\noalign{\smallskip}
\end{tabular}
\\
$^{\dag}$ The two spectral indices are between 235 - 610 and 610 - 1500 MHz, 
respectively.
\end{table*}
\end{center}

\section{Radio Data Analysis}\label{obs}
We have observed \myclus with the GMRT and the VLA in Dec. 2012 and Jul. 2013, 
respectively. The summary of the observations is 
given in Table ~\ref{obstab}.
\subsection{GMRT}
\myclus was observed with the GMRT  
using the dual frequency mode (610 and 235 MHz). 
This mode allows simultaneous recording of data at 610 and 235 MHz 
with one polarization at each frequency.
The data reduction was carried out using NRAO Astronomical Image Processing 
System
 (AIPS). The data at 610 and 235 MHz were saved in separate FITS files and the 
standard steps of data reduction such as flagging (removal of bad data), calibration and 
imaging and self-calibration were followed. 
The standard calibrators 3C48 and 3C147 were used for flux and bandpass calibration and  
the sources 0409-179 and 0521+166 were used for phase calibration.
The calibrated target source data were examined and bad data were removed.
 The final visibilities on the target were then split from the multi-source file and 
averaged in frequency appropriately to avoid being affected by bandwidth smearing and 
at the same time to reduce the size of the data.
The frequency averaged data were then imaged in \textsc{AIPS}. 
A few iterations of phase 
self-calibration and one iteration of amplitude and phase self-calibration were 
carried out. At an intermediate stage in self-calibration, when most of the 
sources were cleaned, the 
clean components were subtracted from the visibilities. The residuals were clipped
 to excise the low level RFI and the clean components were added back. These visibilities 
were used for further iterations of self-calibration. The final images were produced 
using robust ``0'' weights and the rms in the same has been reported in Table ~\ref{obstab}. 
Restriction of uv-range was used to produce images with lower resolutions for the analysis of extended emission.
The images were corrected for the GMRT primary beam using the polynomial approximation 
to the beam using the task `PBCOR'. A conservative estimate of amplitude 
calibration error ($\sigma_{\mathrm{amp}}$) 
of $10$ per cent at both 235 and 610 MHz was considered in calculating the 
errors on the measured flux densities 
of the sources. The \textsc{AIPS} task \textsc{TVSTAT} was used for 
measuring the total flux density of extended sources  
and the task \textsc{JMFIT} was used for point sources. The error on 
the flux density ($\Delta S_{\nu}$) of a source of strength 
$S_{\nu}$ at a frequency $\nu$, 
that has an extent of $N_{b}$, number of beams, is calculated as 
$\Delta S_{\nu}= [\,(\sigma_{\mathrm{amp}}*S_{\nu})^2 + 
(\sigma_{\mathrm{rms}}*\sqrt{N_{b}})^2]^{1/2}\,$, 
where $\sigma_{\mathrm{rms}}$ is the rms noise in the image.

\subsection{VLA}
The VLA in C-configuration was used to observe \myclus in the frequency band $1 
- 2$ GHz. The 1 GHz 
bandwidth contained 16 frequency sub-bands, each with 64 channels.
The pre-processed data provided by NRAO were downloaded and processed further in 
Common Astronomy Software Applications 
(\textsc{CASA}). 
The pre-processing involved flagging and calibration.
 Manual flagging was carried out to remove the residual bad data. From the 16 subbands 
five (1, 2, 3, 8, 9) were flagged. Two channels at the edges 
of each of the sub-bands were removed and 10 channels were averaged to a single channel.
The data were averaged to 10s to reduce the data volume. The final visibilities with 
6 channels per sub-band and 10s samples were used for imaging.
The \textsc{CASA} task `clean', with the multi-scale multi-frequency synthesis 
(MS-MFS) imaging,  
was used. Three iterations of phase self-calibration were carried out. The 
rms in the image produced using 
robust = 0 (weighting=`Briggs') for the visibilities is reported in 
Table~\ref{obstab}.
The task `widebandpbcor' was used to correct for the effect of the primary beam 
of the VLA antennas.
 An amplitude calibration error of $10$ per cent has been considered in 
calculating the errors on the measured 
flux densities.

\section{Radio Results}\label{images}
The GMRT 610 MHz image is presented in Fig.~\ref{fig1} along with an overlay on 
the Digitized Sky Survey R-band image. The GMRT 235 MHz and the VLA 1500 MHz 
images are presented in Fig.~\ref{fig2}. The region of \myclus consists of a 
 few discrete sources and a large elongated source.
The radio properties of these are reported in the following 
sub-sections.

\subsection{Discrete radio sources}\label{srcs}
\begin{figure}
 \includegraphics[trim=1.0cm 6.0cm 2cm 7cm,clip,height=6.4cm]{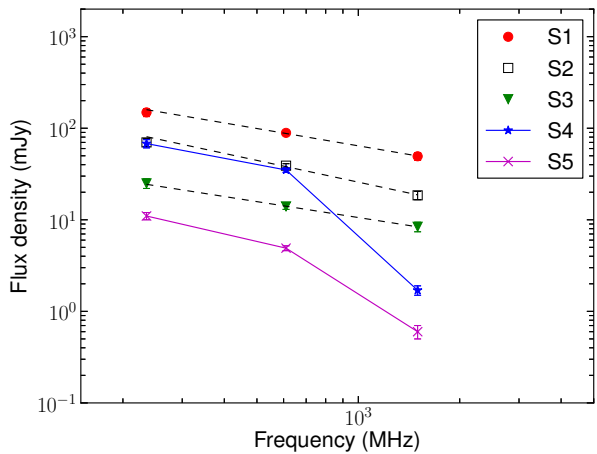}
 \caption{Integrated spectra of the sources S1 -- S5 in PLCKG200.9-28.2. The 
black dashed lines show the best fit spectra 
 for S1, S2 and S3. The solid lines join the flux density points for S4 and S5 
separately.}
\label{srcspec}
\end{figure}

The discrete radio sources in the central region of the \myclus field are 
named S1 -- S5 and those near the large elongated radio source as A and B.
The discrete sources S1-S5 are all resolved at 
610 MHz. The sources S1, S2 and S3 have optical counterparts.
Sources S4 and S5 are both diffuse and have no optical counterparts in the DSS 
image.
The integrated spectral indices of S1, S2, and S3 are in the range 0.6 - 0.8 
(Fig.~\ref{srcspec}, Table ~\ref{srctab}). The diffuse sources S4 and S5 show 
steepening in their 
spectra at higher frequencies. S4 and S5 may be radio 
lobes associated with S3.
Towards the south of the cluster, embedded in the extended emission is the 
source A that can be identified with the galaxy 2MASX J$04502353-0302367$ 
(Fig.~\ref{gmrtopt}). 
The unresolved source B lies beyond the edge of diffuse source towards the  
southwest, has no optical 
counterpart and has an inverted spectrum (Table~\ref{srctab}).

\subsection{A radio relic}
There is an elongated, diffuse radio source located at a 
projected distance of about 0.9 Mpc towards the south-west of the X-ray 
position of the cluster \myclus (PC12, Fig.~\ref{gmrt610}). 
 The discrete sources A and B in the region show no  
morphological connection like jets to the extended source at the 
highest resolution of $\sim4''$ at 610 MHz (in an image with uniform weights 
for the visibilities, not shown). We refer to this source as the radio relic.
Assuming the relic to be at the redshift of the host cluster, the largest 
linear extent of the relic is 1020 kpc and the largest width is 280 
kpc. The properties of the relic are summarised in Table ~\ref{relicprop}. 

 \begin{figure}
 \includegraphics[trim=1.0cm 6.0cm 2cm 7cm,clip,height=6.4cm]{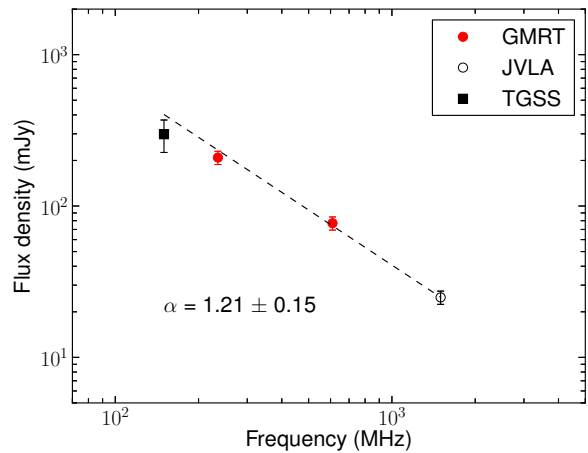}
 \caption{Integrated spectrum of the relic in PLCKG200.9-28.2. The dashed line 
shows the best fit to 
 the spectrum. } \label{intspec}
\end{figure}
 
\subsubsection{Integrated spectrum}
The flux density of the relic was measured in the each of the 235, 610, 
and 
1500 MHz image using identical regions.
The flux density in the relic was 
measured in a region covering the extent of the relic which also included 
the source A. The flux density of A (Gaussian fit), which is the only point 
source in the region, was subtracted from it.
The resulting flux densities of the relic and the spectral indices are reported in 
Table~\ref{relicprop}. The 150 MHz image of the region of \myclus 
 available from the TIFR GMRT Sky Survey 
(TGSS\footnote{http://tgss.ncra.tifr.res.in/}, PI S. Sirothia) was also used.
The rms noise in the image is $5$ \mjyb (beam is $\sim25''$). The flux density 
of the relic in this image, assuming 
a $20$ per cent error, is reported (Table~\ref{relicprop}, 
Fig.~\ref{intspec}).
The integrated spectrum of the relic is a single power-law with a spectral 
index $1.21\pm0.15$.

\begin{figure*}
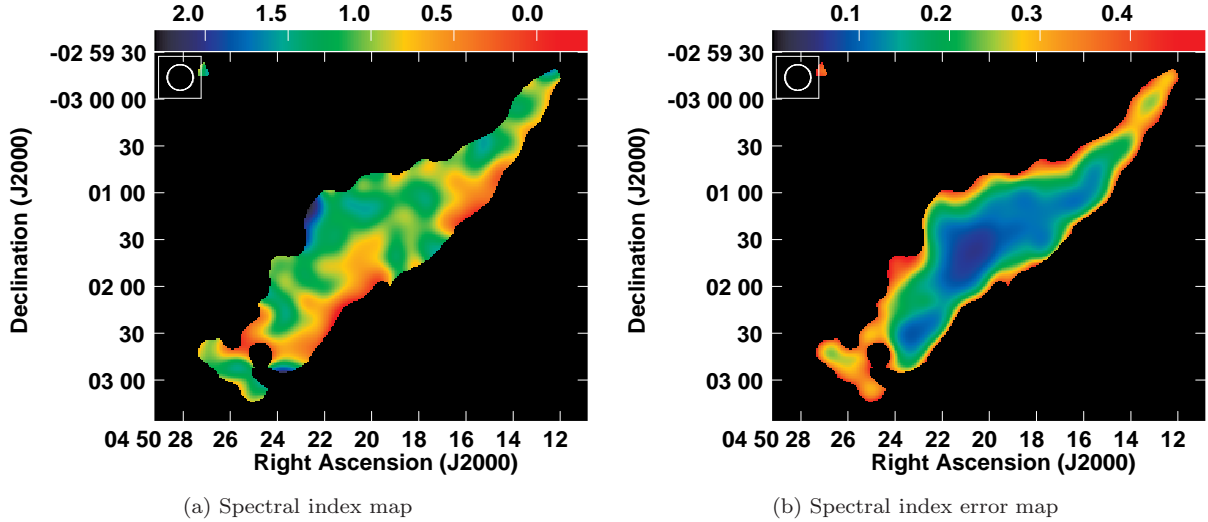

\subfloat[Spectral index map]{\includegraphics[trim=0.0cm 0.0cm 0cm 
0cm,clip,height=6.5cm]{SPIX_MAP.PS}}
\subfloat[Spectral index error map]{\includegraphics[trim=0.0cm 0.0cm 0cm 
0cm,clip,height=6.5cm]{SPIX_MAP_N.PS}}
\caption{ (a) -- The spectral index map of the radio relic between 
235 and 610 MHz. (b) -- The 
corresponding noise map. The resolution of the maps shown in upper left 
corner of each panel is $16.4''\times16.4''$. 
\label{spec26}
}
\end{figure*}

\subsubsection{Spectral index map}
With radio observations at two or more frequencies with overlapping 
uv-coverages, spectral index maps of an extended source can be constructed.
The steps to make the spectral index map between 235 and 610 MHz are described briefly.
The images at each frequency were made using uniform weights for the 
visibilities in the overlapping uv-range of 0.2 - 19 k$\lambda$. These were 
corrected for the effect of the respective primary beams and 
convolved to a common resolution of $16.4''\times16.4''$. 
The rms noise ($\sigma_{\mathrm{rms}}$) in the resulting 
images were 0.16 and 0.48 \mjyb at 610 and 235 MHz respectively. 
The regions of the image with flux densities below $3\sigma_{\mathrm{rms}}$ 
were blanked. The resulting images were used in the task `COMB' to make spectral 
index maps.
The spectral index and the corresponding noise map are presented 
in Fig.~\ref{spec26}. 
The spectral indices at the south western edge (outer edge) of the relic are
flatter ($\sim 0.7$) and steepen towards the north east ($0.9 - 1.3$).

\begin{table}
\centering
\caption[]{\label{relicprop}Properties of the radio relic.}
\begin{tabular}{ll}
\hline\noalign{\smallskip}
RA$_{\rm J2000}^{\ddag}$ & 04h50m20.7s\\
\smallskip
DEC$_{\rm J2000}$ &-03$^{\circ}$01$^\prime$37.6$^{\prime\prime}$ \\
\smallskip
LLS$^{\dag}$ & 1020 kpc\\
\smallskip
Largest Width  & 280 kpc\\
\smallskip
$S_{150 \mathrm{MHz}}$ & $298\pm82$ mJy\\
\smallskip
$S_{235 \mathrm{MHz}}$ & $209\pm21$ mJy\\
\smallskip
$S_{610 \mathrm{MHz}}$ & $77\pm 8$ mJy\\
\smallskip
$S_{1500 \mathrm{MHz}}$ & $24.9\pm2.5$ mJy\\ 
\smallskip
$\alpha$ & $1.21\pm0.15$\\
\smallskip
$P_{1.4\mathrm{GHz}}^{*}$  & $ (3.9\pm0.4) \times10^{24}$ W Hz$^{-1}$\\
Mach number, $M_{int}$ & $2.8\pm0.5$\\
\smallskip
Mach number, $M_{edge}$ &$3.3\pm1.8$ \\
\noalign{\smallskip}
\hline\noalign{\smallskip}
\end{tabular}
\\
$^{\ddag}$ RA$_{\rm J2000}$ and DEC$_{\rm
  J2000}$ are the coordinates of the peak at 610 MHz. $^{\dag}$ Largest 
linear size. $^*$ Power at 1.4 GHz is calculated by extrapolating the 
$S_{1500\mathrm{MHz}}$
to 1.4 GHz using the spectral index, $\alpha$.
\end{table}

\section{X-ray analysis}\label{xrayobs}
\subsection{Data analysis}
\myclus was observed for 17~ks with \xmm in February 2011 
(ObsID 0658200801). 
The EPIC data were reduced with {\tt SAS} version 
11.0.0\footnote{http://heasarc.gsfc.nasa.gov/docs/xmm/xmmhp\_xmmesas.html}, 
primarily using the Extended 
Source Analysis Software 
(XMM-ESAS) subpackage.
We follow the analysis method introduced by \citet{2008A&A...478..615S} for 
EPIC MOS. For EPIC pn, the methods are described in 
\citep[e.g.,][]{2012ApJ...747...32B, gia13}. A brief summary of the 
analysis is presented here.
A filtered event list is created by removing periods of 
excessive count
rate in the light curve caused by proton flaring events (retaining 16.1~ks,
15.7~ks, and 4.8~ks of the exposures for MOS1, MOS2, and pn, respectively), 
excluding non-functioning MOS
CCDs (in this case CCDs 4, 5, and 6 for MOS1), and identifying and excluding
point sources.
Unfortunately, all data were highly contaminated by proton flaring, so we
exclude periods with high incidence of flares and keep those with only lower level flaring.

Spectra and images are then extracted for source regions and the quiescent particle
background, the latter of which are derived from a database of filter wheel closed observations
that are matched to data using the unexposed corners of the chips.
Remaining background components (residual soft proton contamination,
instrumental lines, 
solar wind charge exchange (SWCX) emission, and the cosmic fore- and backgrounds
from the local hot bubble, the Galaxy, and extragalactic unresolved point sources)
are all explicitly modeled and determined empirically during spectral fits in {\tt XSPEC}.
While robust best-fits of these components can be trickier to obtain in short
exposures such as these, the small angular extent of the cluster 
leaves much of the FOV available for background modeling.
We fix the absorbing column to the Galactic value ($n_H = 3.9 \times 10^{20}$~cm$^{-2}$)
and the redshift of the cluster ($z = 0.22$) to that reported previously in PC12.
The local hot bubble (unabsorbed $kT = 0.1$~keV) and Galactic emission
(absorbed, two temperature model with $kT$s of 0.1~keV and 0.3~keV) are further
constrained by a ROSAT all sky survey spectrum from an annular region with
radius 1~deg$^2$ centered on the cluster position, obtained with the online
X-ray background tool\footnote{http://heasarc.gsfc.nasa.gov/cgi-bin/Tools/xraybg/xraybg.pl}.
Due to the high level of flaring, the soft proton component is modeled as a broken
power law with indices $\Gamma$ and break energies $E_{\rm b}$ left free, leading
to consistent values between all instruments with hard ($\Gamma \sim 0.3$) indices below
$E_{\rm b} \sim 1.5$~keV and softer ($\Gamma \sim 1.0$) indices above $E_{\rm b}$.
Also, no SWCX lines are evident in any of the spectra.

For image analysis, we combine the data, background, and properly scaled
exposure maps (in equivalent MOS2 medium filter count rates) from all the three 
instruments.
Separate background images include the contributions of the non-X-ray particle background
(scaled and taken from stowed exposures), the soft proton background (scaled and
taken from the ESAS CALDB), and the cosmic X-ray background, of both diffuse and 
unmasked point source emission, as modeled during
spectral fitting (convolved with exposure maps).
When extracting a profile, the size of a bin is increased until 50 net (data minus background)
source counts are accumulated, which is then divided by the average exposure within
the bin and area excluding masked pixels.

\subsection{X-ray Results}\label{xrayresults}
\begin{figure*}

\subfloat[\xmm 0.4 - 1.4 keV with 610 
MHz contours.]{\includegraphics[trim=0.5cm 
0.3cm 1.8cm 1.4cm,clip,height=8.0cm]{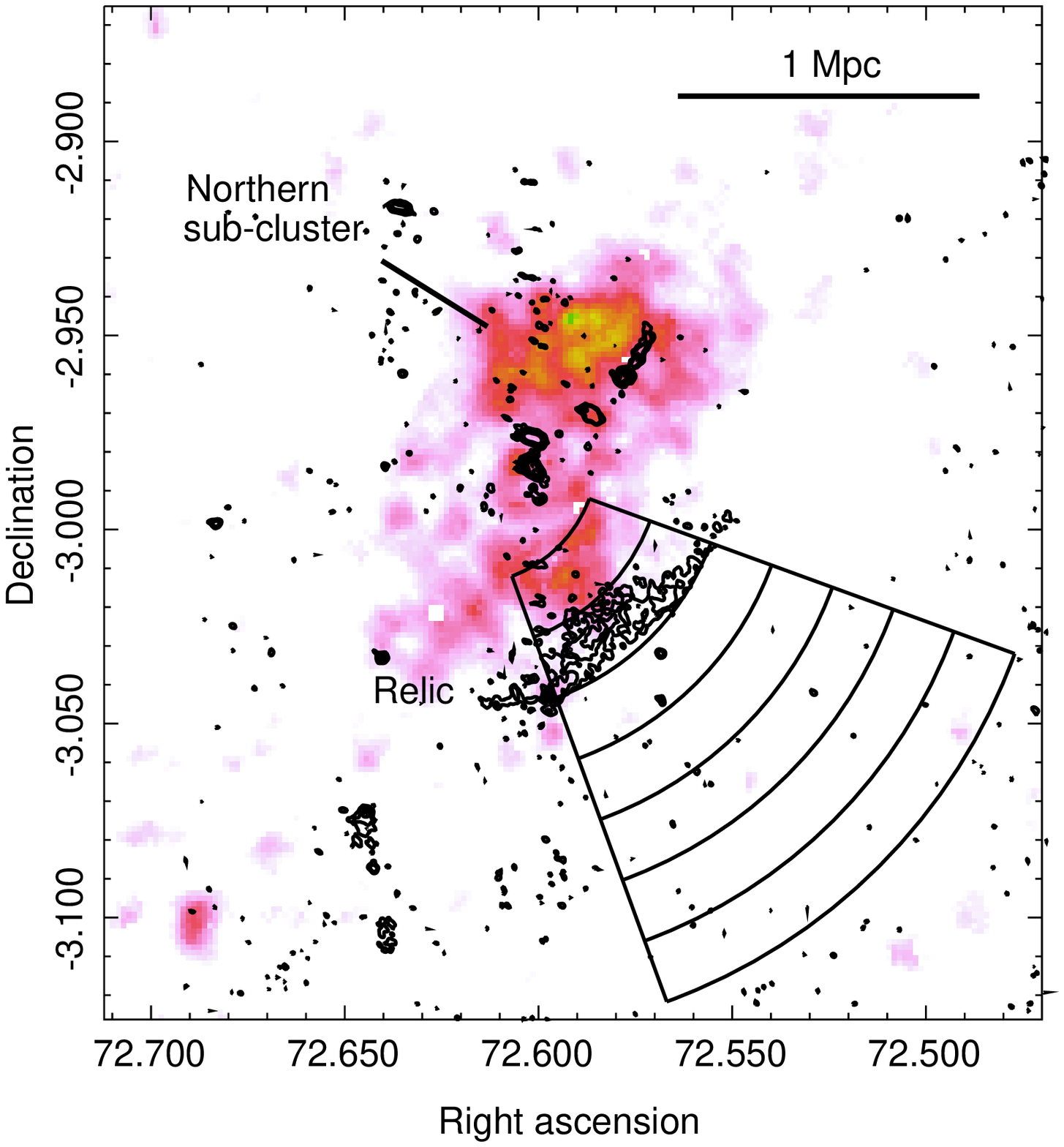}\label{xrayb}}
\subfloat[Surface brightness profile.]{\includegraphics[trim=0.2cm 
0.1cm 0.0cm 0.0cm,clip,height=8.0cm]{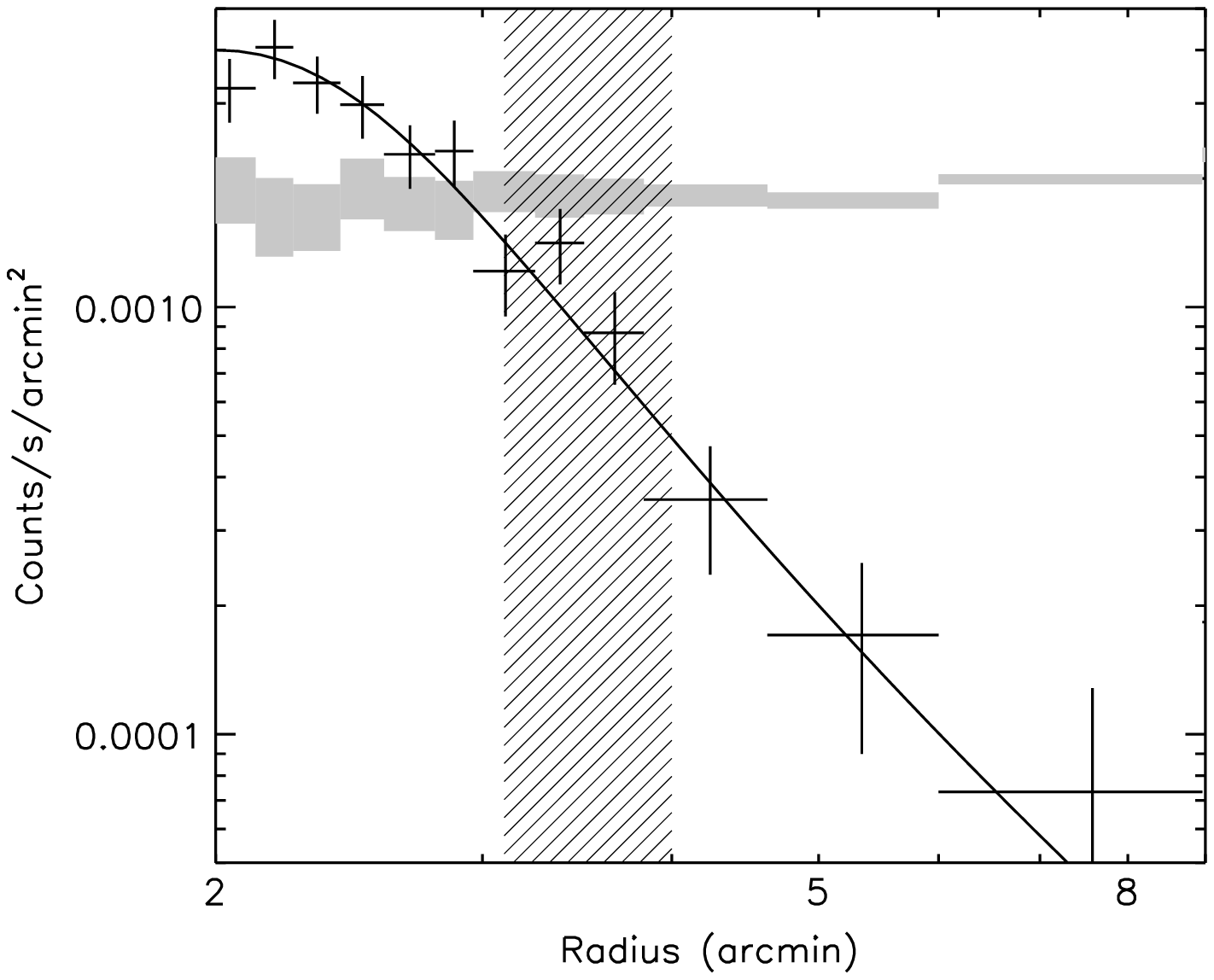}\label{profile}}
\caption{(a)-- \xmm 0.4-1.4 keV adaptively smoothed, greyscale image with 
overlays of the radio relic (black contours same as Fig.~\ref{gmrtopt}) and of 
the circular sector used to extract the X-Ray surface brightness profile in the 
same energy band.  
Each arc corresponds to a width of $1'$.  
(b)-- 0.4-1.4 keV surface brightness profile (crosses, 1-sigma errors shown).  
The hatched region indicates the location of the relic and the gray bars show 
the estimated background level in each bin.  
The solid line shows a beta profile with no density jump.
Because the \xmm data are not sufficiently sensitive to detect the 
emission beyond the relic, a density jump cannot be ruled out. \label{xray}}
\end{figure*}

The \xmm image in 0.4 - 1.4 keV band is presented in Fig.~\ref{xrayb}. 
Due to the residual soft proton contamination, we restrict our spatial analysis 
to this softer band to maximize signal-to-noise; however, note that all spectra 
are fit over the full 0.3-12 keV band so that temperatures are well constrained.
The morphology in the image can be divided into northern and southern 
sub-clusters. 
The relic is at the south-western periphery skirting the southern sub-cluster.
We find temperatures of $5.3^{+1.0}_{-0.9}$~keV and $6.1^{+1.9}_{-1.2}$~keV 
for the northern and southern sub-clusters, respectively.
While the southern subcluster temperature may be biased high by a shock related 
to the radio relic, and/or the northern subcluster temperature may be biased low 
due to a cool core, additional X-ray observations are required to say more.
We measure the total average temperature of the entire cluster to be 
$5.6 \pm 0.8$~keV, with an unabsorbed luminosity of 
$L_{X[0.1 - 2.4 \mathrm{keV}]} = (1.05 \pm 0.05) \times 10^{44}$~erg s$^{-1}$.
This is somewhat hotter than the value found by PC12 of $4.5 \pm 0.7$~keV, 
likely a consequence of the high background and difficulty in modeling it accurately. 
This mild discrepancy should be taken as a demonstration of the impact of systematic uncertainties.
Our slightly higher value of $L_X$ is likely due to including the southernmost emission
from the cluster, which falls outside the $R_{500}$ radius considered by PC12.
The abundance is constrained to be no higher than 0.38 relative to solar, consistent
with clusters generally.

In Fig.~\ref{xrayb}, we extract a surface brightness profile within a circular 
wedge that roughly aligns with the total curvature of the relic.
The Fig.~\ref{profile} shows the surface brightness profile along with the 
total background
level (solid gray boxes, indicating 1$\sigma$ uncertainties) and the location of the
radio relic (hatched region).
The edge of the relic falls well below the background level, making any estimate of
a surface brightness jump impossible, although there is a large 
drop in surface brightness just outside the relic.
Since the pre-shock region is deep in the background, we cannot formally 
constrain the density jump, and hence estimate Mach number from these data. 
We model a density jump in the surface brightness profile assuming a power 
law density profile ($\propto n_e^\Gamma$) with a discontinuity fixed at the 
location of the edge of the relic.
If we fix the index of the density profile to $\Gamma=1.7$, the 
best-fit value 
on both sides of the potential shock, the density jump is constrained to be less 
than a factor of 2 ($90$ per cent confidence), or $M<1.8$.  Whether the X-ray 
estimate 
of the Mach number is actually in tension with the radio estimate, however, 
will be confirmed by deeper observations.
Assuming no jump at the radio relic location, the profile can be described by a 
classic $\beta$-model with core radius of $1.0^{+0.9}_{-0.4}$~arcmin,
$\beta = 0.6^{+0.4}_{-0.1}$, and central surface brightness of 
$(3.9^{+1.0}_{-0.7}) \times 10^{-3}$~counts s$^{-1}$ arcmin$^{-2}$.

\section{Discussion}\label{discussion}
\subsection{Relic power, size and cluster mass}\label{mass}
\begin{figure*}
 \includegraphics[trim=2.4cm 2.8cm 1.5cm 
4cm,clip,height=11cm,angle=-90]{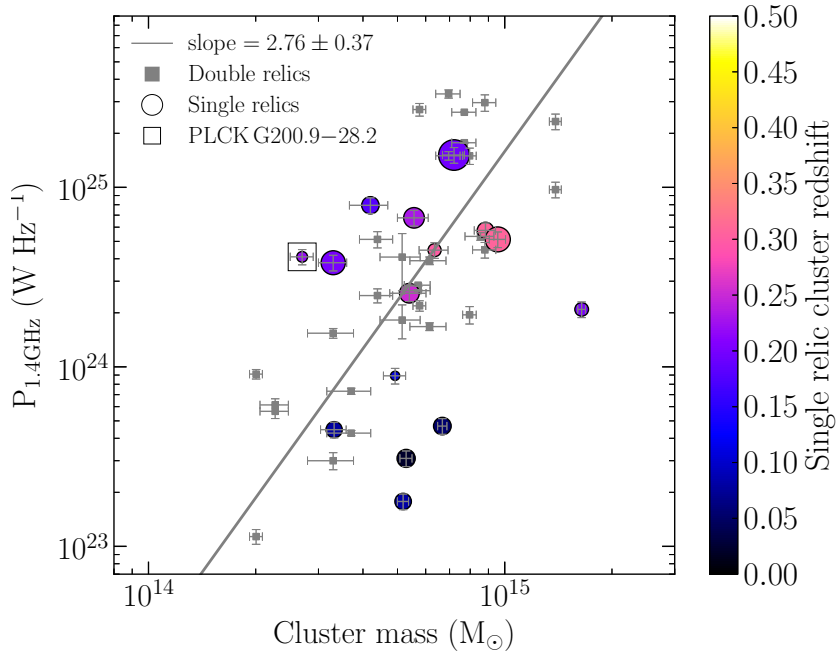}
 \caption{The 1.4 GHz power of radio relics is plotted against the host cluster 
mass. The circles show the single relics, including \myclus. The 
sizes of the circles are scaled 
according to the largest linear sizes of the relics and the colour denotes the 
redshift. \myclus relic is highlighted with an empty square. The grey squares 
show the double 
relics. The best fitting scaling relation for the double relics is shown by the 
dashed line. The data for the known relics 
from \citet{gas14,gas15,gas17} and the best fitting slope have been revised and 
plotted (Sec.~\ref{mass}).
\myclus has the lowest mass among the hosts of single relics. 
The \myclus relic has the smallest size among the single relics with radio 
powers greater than $10^{24}$ W Hz$^{-1}$.  
} 
\label{powermass}
\end{figure*}

The correlation between the 
relic power at 1.4 GHz ($P_{1.4 \mathrm{GHz}}$) and the host cluster mass 
 has been reported for the sample of known single and double radio relics 
\citep{gas14}. 
The \myclus relic plotted in the $P_{\mathrm 
1.4 \mathrm{GHz}}$ - mass plane along with other known radio relics is shown in 
Fig.~\ref{powermass}. The literature data presented in \citet{gas14} have 
been revised for relics with unknown spectral indices 
\footnote{The spectral index of -1.0 was mistakenly used in \citet{gas14} to 
calculate the radio powers of the relics with unknown spectral indices. We have 
used the correct value of 1.0 and revised the radio powers.} and updated to 
include the radio relics in the clusters A3527-bis and PSZ1 G108.18-11.53 
\citep{gas15,gas17}. The revised scaling 
relation has a slope of $2.76\pm0.37$ and an intercept of $-16.2\pm5.3$ 
(Fig.~\ref{powermass}).

\myclus has the lowest mass among the clusters with single relics 
and is the third lowest mass cluster among all the radio relic clusters. 
The two clusters with masses less than \myclus are Abell 3365 \citep{wee11b} 
and Abell 3376 \citep{bag06,kal12}, both with double relics. As 
compared to the predictions of the 
simulations by \citet{nuza12}, the probability of finding a radio relic 
in a cluster of mass $\sim3.3\times10^{14}$ M$_{\odot}$ and redshift $\sim 0.2$ 
is $<$few per cent with the current instruments \citep{gas14}. The case of the 
\myclus relic provides an example of an  
 even lower mass cluster with a powerful radio relic.
The sample of single relics does not follow the radio power-mass scaling 
 and 
\myclus is a factor of 10 more powerful in radio than expected from the 
correlation.

We investigated whether the largest linear sizes of the single relics show 
scaling with radio power, mass or redshift. In Fig.~\ref{powermass}, the sizes 
of the circles denoting single relics are scaled according to their largest 
linear sizes. We point out that among the powerful 
($>10^{24}$ W 
Hz$^{-1}$) single radio relics, \myclus has the smallest size. However, there 
is no trend in size with respect to the host cluster mass, redshift and relic 
radio power. The effects of projection may be the important factors in the 
apparent sizes of the relics.

\subsection{A merger shock ?}\label{merger}
The cluster \myclus in X-rays shows a disturbed morphology indicating an ongoing merger. 
There is a northern clump of gas that extends with a lower brightness emission to the south 
reaching the location of the relic (Fig.~\ref{xray}). The spectral index map of the relic shows flatter spectrum 
emission towards the outer edge and gradual steepening towards the inner side strongly indicating 
 an underlying merger shock.
Under the assumptions of DSA, if $\alpha_{\mathrm{inj}}$ is the spectral index 
at injection then the Mach number of the shock is given by 
\citep[e.g.][]{bla87},
\begin{equation}
 M = \sqrt{\frac{2\alpha_{\mathrm{inj}} +3}{2\alpha_{\mathrm{inj}}-1}}.
\end{equation}
In the case of \myclus relic, the $\alpha_{\mathrm{inj}}$ is the spectral index 
at the outer edge of the relic, equal to $0.7\pm0.2$ (Fig.~\ref{spec26}). This 
implies, $M=3.3\pm1.8$ 
for the relic. In a scenario where there is continuous 
injection at the shock, a spectral index of $\alpha_{\mathrm{inj}}+0.5$ is 
expected. 
The integrated spectral index of $1.21\pm0.15$ is consistent with this 
expectation and implies $M=3.2\pm1.3$. This relic is one of the cases where 
the integrated spectra and the injection spectral index are consistent with 
continuous injection scenario. 

If a shock is detected in X-rays, then a Mach number independent of the 
detection in radio can be estimated. The X-ray derived Mach numbers have been 
found consistent with those from radio spectra in some relics \citep{aka13} but 
may also have a mismatch \citep{Ogrean13}.
From the \xmm X-ray data on \myclus, a
Mach number of $<$1.8 is expected presuming a density jump, which is less than 
that from the radio spectral indices. Our recently approved {\emph {Chandra}}
observations will provide precise measurements of the Mach number.

The morphology of \myclus relic is overall arc-like with a notch at the outer 
edge (Fig.~\ref{fig1}). The notch is a low brightness region which breaks the 
smoothness of the outer edge of the relic. It is prominent in higher frequency 
images as compared to that at 235 MHz. Such features can indicate 
the changes in the underlying physical conditions in the shock. It can also be 
an effect of a density change due to an infalling structure such as proposed in 
some simulations \citep[e.g.][]{pau11}.

\subsection{X-ray and SZ position offset}\label{offset}
Cluster mergers can stir the ICM and lead to complex distributions of 
density and temperatures. The X-ray surface brightness traces regions of high 
electron densities ($\propto n_e^{2}$) and the SZ is sensitive to the pressure 
($\propto n_e T$) along the line of sight. An offset in the peaks of these 
signals can be used as an indicator of the density and temperature distribution 
in the disturbed ICM. There are examples of merging clusters that show presence 
of X-ray -SZ offsets, such as Abell 2146 \citep{rod11}
and Bullet cluster \citep{malu10}. 
Simulations have shown that the offset is sensitive to initial relative velocity 
of the merging clusters and the mass ratio \citep[e. g.][]{molnar12,Zhang14}.
The cluster \myclus discussed in this work has its X-ray peak and the Planck 
detection 
peak offset by $3.4'$, which is the extreme in the \plck sample (PC12). 
The \plck SZ positions have a mean and median errors of $1.5$ and $1.3'$
respectively \citep{2013A&A...550A.130P}. The X-ray peak is located at the 
northern sub-cluster (Fig.~\ref{gmrt610} and ~\ref{xrayb}) and the \plck SZ 
position is separated from it by $3.4'$ (700 kpc) in the direction of the radio 
relic. Due to the presence of a shock at the relic, the region is expected to 
be overpressured and thus may result in shifting the peak of the SZ-signal.
The offset in the direction of the relic indicates possible 
physical origin for the offset in addition to the position reconstruction 
uncertainty of \plck. Based on the results of simulations, the offset can 
be explained as a result of two comparable mass sub-clusters with mass ratio 
between 1 and 3 \citep{Zhang14}. Deep optical observations tracing the galaxy 
distribution in this cluster will be useful to measure the mass ratios of the 
sub-cluster masses. Due to the large uncertainty in the \plck position we 
cannot analyse the offsets for a statistical sample of merging clusters. 
 However, the offset in the X-ray and SZ positions opens an additional 
probe for understanding the properties of merging galaxy clusters.

\section{Conclusions}\label{conc}
We have presented the discovery of a single arc-like relic in the  galaxy 
cluster \myclus discovered by the \plck satellite. This cluster has the lowest 
mass among the clusters known to host single arc-like radio relics.  The main 
results and conclusions from our work are as follows:
\begin{enumerate}
 \item The GMRT 235 and 610 MHz and VLA 1500 MHz images confirm the presence 
of a radio relic of size $\sim 1 \times 0.28$ Mpc located at a projected 
distance of 0.9 Mpc from the centre of the cluster PLCKG200.9-28.2.
\item The \xmm X-ray images show merging northern and southern sub-clusters in 
this cluster and the relic is located at the south-western periphery of the 
cluster.
\item The radio relic has an integrated spectral index of $1.21\pm0.15$. The 
235 - 610 MHz spectral index map shows steepening from outer to inner edge of 
the relic. This is consistent with expectations for a relic generated by 
an underlying merger shock.
\item The radio relic has a 1.4 GHz power of $(3.9\pm0.4) \times10^{24}$ W 
Hz$^{-1}$. The \myclus is the lowest mass single relic cluster known.
The radio relic is the smallest in size in the sample of single relics with 
radio power $>10^{24}$ W Hz$^{-1}$.
\item The analysis of the X-ray brightness profile along relic shows the 
possibility of a putative shock. Future deep {\emph Chandra} observations 
will be able to characterize the shock in X-rays.
\item The cluster \myclus has the extreme offset of $3.4'$ (700 kpc) in the 
positions of the peak in X-ray and in \plck SZ. The \plck position is offset 
toward the position of the relic where high pressure region may be expected due 
to merger shock. The X-ray - SZ position offset may be of physical origin 
and an important probe of merging cluster with future better measurements in 
SZ positions.
\end{enumerate}

\section*{Acknowledgements}
We thank the anonymous referee for comments that helped to improve the paper.
RK acknowledges the support from DST-INSPIRE Faculty Award from the Department 
of Science and Technology, Government of India. 
GB, TV, RC acknowledge partial support from PRIN-INAF 2014. 
F.d.G. is supported by the VENI research programme with project number 
1808, which is financed by the Netherlands Organisation for Scientific Research 
(NWO). We thank the staff of the GMRT that made these observations 
possible.  GMRT is run by the National Centre for Radio 
Astrophysics of the Tata Institute of Fundamental Research.
The National Radio Astronomy Observatory is a facility of the 
National Science Foundation operated under cooperative 
agreement by Associated Universities, Inc.
Based on observations obtained with {\emph XMM-Newton}, an ESA science 
mission with instruments and contributions directly funded by 
ESA Member States and NASA. Basic research in radio astronomy at the
Naval Research Laboratory is supported by 6.1 base funding.
This research has made use of the NASA/IPAC Extragalactic Database (NED) which 
is operated by the Jet Propulsion Laboratory, California Institute of 
Technology, under contract with the National Aeronautics and Space 
Administration.

\bibliographystyle{mnras}
\bibliography{mybib_1}




\bsp	
\label{lastpage}
\end{document}